# ARABIC TEXT MINING

**Sumaia Mohammed AL-Ghuribi [a,b*] and Shahrul Azman Mohd Noah [b]**

[a] Department of Computer Science, Faculty of Applied Sciences, Taiz University, Taiz 6803, YEMEN
[b] Center for Artificial Intelligent Technology, Faculty of Information Science and Technology, Universiti Kebangsaan Malaysia, Bandar Baru Bangi 43600, Selangor, MALAYSIA

## Abstract

The rapid growth of the internet has increased the number of online texts. This led to the rapid growth of the number of online texts in the Arabic language. The enormous amount of text must be organized into classes to make the analysis process and text retrieval easier. Text classification is, therefore, a key component of text mining. There are numerous systems and approaches for categorizing literature in English, European "French, German, Spanish", and Asian "Chinese, Japanese". In contrast, there are relatively few studies on categorizing Arabic literature due to the difficulty of the Arabic language. In this work, a brief explanation of key ideas relevant to Arabic text mining are introduced then a new classification system for the Arabic language is presented using light stemming and Classifier Naive Bayesian (CNB). Texts from two classes—politics and sports—are included in our corpus. Some texts are added to the system, and the system correctly classified them, demonstrating the system's effectiveness.

## 1. Introduction

Text mining is a field that attempts to glean meaningful information from natural language text. It is used to discover new, previously unknown information by automatically extracting information from different written resources. Most previous studies of data mining have focused on structured data. However, in reality, substantial portions of the available information are stored in text databases that consist of a large collection of documents from various resources. Compared to data mining, which aims to find patterns in structured data, text mining is more challenging as the analyzed data is unstructured.

Most data and text mining studies have been conducted in English and other European languages. There is little ongoing research on Arab text mining in the Arabic language, while a huge amount of Arabic online information appears in collections of unstructured text. As a result, Arabic text mining becomes necessary to make sense of all this information and to discover knowledge from the collection of unstructured Arabic text.

Research into Arabic has been severely hampered by the lack of an adequate lexical database that would provide information about its distributional and structural characteristics [1]. The main reason is the complex and rich nature of Arabic. The Arabic language consists of 28 letters. It is written from right to left. It is a very complex morphology, and most words have a tri-letter root. The rest have either a quad-letter root, penta-letter root or hexa-letter root.

---

[*] Corresponding author.
 Email addresses: somaiya.ghoraibi@gmail.com (SM AL-Ghuribi), shahrul@ukm.edu.my (SA Mohd Noah).

The remainder of the paper is organized as follows. Section 2 contains some explanation for some important points related to our work. Section 3 presents some characteristics of the Arabic language that make it complex. Classification methods for Arabic documents are in section 4. Section 5 presents a new system for Arabic text mining. Finally, the conclusion and future works are given in section 6.

## 2. Some major points in this search:

### 2.1 Data Mining: It is a field in computer science and can be defined as follow:

- Data mining is extracting patterns from large data sets by combining statistics and artificial intelligence methods with database management. Or in other words, the exploration and analysis of large quantities of data in order to discover valid, novel, potentially useful, and ultimately understandable patterns in data.

  Valid: The patterns hold in general.

  Novel: We did not know the pattern beforehand.

  Useful: We can devise actions from the patterns.

  Understandable: We can interpret and comprehend the patterns.

Most previous studies of data mining have focused on structured data. However, in reality, substantial portions of the available information are stored in text databases consisting of large documents from various resources. Approximately 90% of the world's data is held in unstructured formats. Here, the idea of text mining, which is concerned with the analysis of unstructured material, is introduced.

### 2.2 Text Mining

Follow are some definitions for text mining:

- Text mining is a field that attempts to glean meaningful information from natural language texts. It is used to discover new, previously unknown information "not found in any individual document" by automatically extracting information from different written resources.

- Text mining is the process of compiling, organizing, and analyzing large document collections to support the delivery of targeted types of information to analysts and decision-makers and to discover relationships between related facts that span wide domains of inquiry.

- Text mining is the discovery or creation of new knowledge from a collection of documents. The new knowledge may be the statistical discovery of new patterns in available data (standard text mining). It may also incorporate AI abilities to interpret patterns and provide more advanced capabilities such as hypothesis suggestion (intelligent text mining). Artificial intelligence, and especially natural language processing, can be used to simulate the human capabilities needed for intelligent text mining.

The difference between data mining and text mining is that in text mining, the patterns are extracted from natural language text rather than from structured databases of facts.

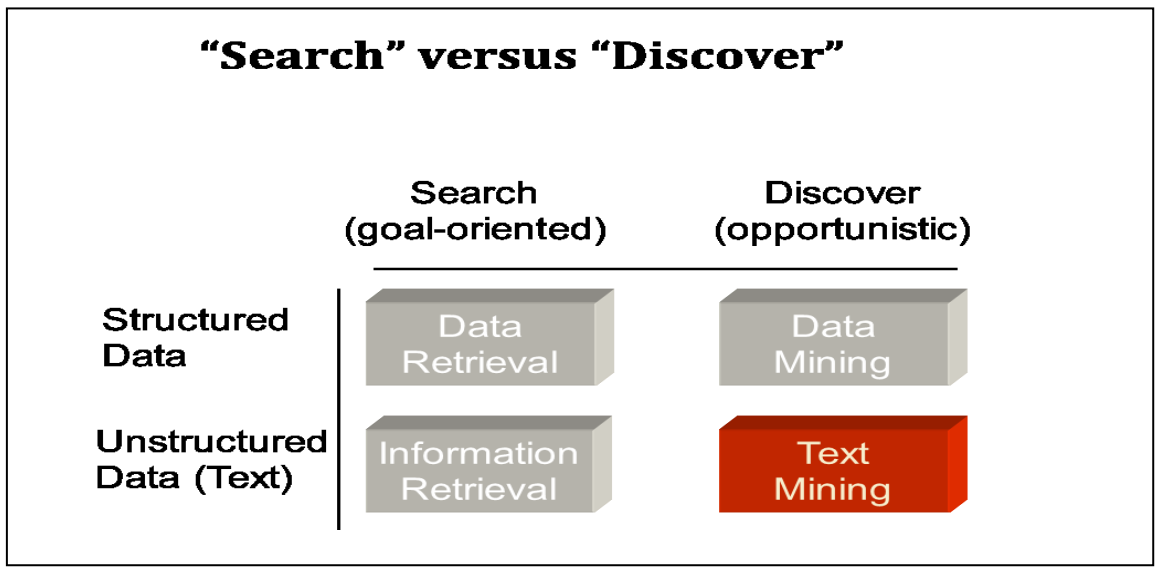

Figure 1: The difference between data mining and text mining

Text mining is different from what we are familiar with in web searches. In search, the user is typically looking for something already known and written by someone else. But in text mining, the goal is to discover unknown information, something that no one yet knows and so not have yet written down.

Due to the vast amount of Arabic online information that appears in collections of unstructured text, Arabic text mining becomes necessary to make sense of all this information and to perform knowledge discovery from collections of unstructured Arabic text. This is the main objective of this work.

### 2.3  Natural Language Processing (NLP)

Natural language processing (NLP) is a field of computer science and linguistics concerned with the interactions between computers and human (natural) languages. In another way, it is a way of translating between computer languages and human languages. Natural language understanding is sometimes referred to as an AI-complete problem because natural-language recognition requires extensive knowledge about the outside world and the ability to manipulate it. The goal of this field is to allow computers to understand what a text says without being given precise values and equations for the text's data. In essence, natural language processing automates the translation process between human and computer languages. While much of this field relies on statistics and models to determine the likely meanings of a phrase, there are and have been many different approaches to this problem. Findings in this field have applications in speech recognition, human language translation, information retrieval, and even artificial intelligence. Natural language processing faces many problems because language is not always consistent, and not all clues to meaning are contained in the language itself. Even a complete account of a language's entire grammar, including all exceptions, does not always allow a computer to parse the information in a text. Some sentences are syntactically ambiguous, words often have more than one meaning, and some combinations of sounds or symbols change their meaning depending on the boundaries of the words — all of which can be problems for a computer that does not understand the context.

### 2.4 Corpus

Corpus (plural, "corpora") is a large and structured set of texts (documents) that have been manually annotated with the correct values. They are employed in statistical analysis and hypothesis tests. It could have texts in numerous languages (a multilingual corpus) or only one language (a monolingual corpus). When we refer to a corpus in this work, we mean a corpus that contains a large set of Arabic documents. There are a lot of Arabic corpora, but most of them are not free. Examples of some corpora are:

- *Al-Hayat Corpus* contains Al-Hayat newspaper articles with value added for Information Retrieval applications development purposes. The data have been distributed into seven subject-specific databases, thus following the Al-Hayat subject tags: General, Car, Computer, News, Economics, Science, and Sport. Mark-up, numbers, special characters and punctuations have been removed. The size of the total file is 268 MB. The dataset contains 18,639,264 distinct tokens in 42,591 articles organized in 7 domains.
- *Quranic Arabic Corpus* is an annotated linguistic resource which shows the Arabic grammar, syntax and morphology for each word in the Holy Quran. The corpus provides three levels of analysis: morphological annotation, a syntactic treebank and a semantic ontology.
- *Arabic Newswire Corpus* is composed of articles from the Agence France Presse (AFP) Arabic Newswire. The source material was tagged using SGML and was transcoded to Unicode (UTF-8). The corpus includes articles from May 13, 1994, to December 20, 2000. The data is in 2,337 compressed (zipped) Arabic text data files. There are 209 Mb of compressed data (869 Mb uncompressed), with approximately 383,872 documents containing 76 million tokens over approximately 666,094 unique words.
- *Arabic Gigaword,* Four distinct sources of Arabic newswire are represented here: Agence France Presse, Al Hayat News Agency, Al Nahar News Agency and Xinhua News Agency. There are 319 files, totalling approximately 1.1GB in compressed form (4348 MB uncompressed and 391619 Keywords).

## 3. Arabic Language Complexity

Arabic is the fifth most popular language in the world. More than 422 million people speak it as a first language, and 250 million speak it as a second language. The Arabic alphabet consists of 28 letters in addition to the Hamza (ء). There is no upper or lower case for Arabic letters like English letters. The letters (أ و ى) are vowels, and the rest are constants. Unlike Latin-based alphabets, the orientation of writing in Arabic is from right to left [2]. It has a very complex morphology, and most words have a tri-letter root. The rest have quad-letter, penta-letter, or hexa-letter roots [3]. Arabic words have two genders, masculine and feminine; three numbers, singular, dual, and plural; and three grammatical cases, nominative, accusative and genitive. A noun has the nominative case when it is the subject, accusative when it is the object of a verb, and genitive when it is the object of a preposition. Words are classified into three main parts of speech, nouns (including adjectives and adverbs), verbs, and particles [4, 5].

- **The complexity of the Arabic Language**

Arabic is a challenging language for a number of reasons, follows are some of them [4]:
1. Orthography with diacritics is less ambiguous and more phonetic in Arabic; certain combinations of characters can be written in different ways.
2. The Arabic language has short vowels which give different pronunciations. Grammatically they are required but omitted in written Arabic texts.
3. Arabic has a very complex morphology as compared to the English language.
4. Synonyms are widespread. Arabic is a highly inflectional and derivational language.
5. Lack of publically freely accessible Arabic Corpora.
6. One word may have more than one meaning in different contexts.
7. In different contexts, one word may have more than a lexical category (noun, verb, adjective, etc.).
8. The form of some Arabic words may change according to their case modes (nominative, accusative or genitive). For instance, the plural of the word (مسافر), which means (traveller) may be in the form (مسافرون) in the case of nominative and the form (مسافرينِ) in the case of accusative/genitive.
9. Moreover, an Arabic word may correspond to several English words. Another example is the Arabic word (وبنفوذها) and its equivalence in English ―and with her influences.

## 4. Arabic Text Classification

Classifying the vast number of texts available on the internet is necessary. Document organization will facilitate analysis and document retrieval. So, one of the critical components of text mining is document classification. Text classification is the process of classifying texts according to their content and placing them in one or more predetermined categories. English documents, European "French, German, Spanish" documents, and Asian "Chinese, Japanese" texts are all classified using a variety of projects, methods, and approaches. However, due to the difficulty of the Arabic language, there are very few studies on Arabic documents [6, 7].

This work will explore a few classification studies of Arabic text. Some studies treat documents as a bag of words, representing the text as a vector of weighted frequencies for each unique word or token. Others discover the root and patterns using different algorithms. Prefixes and suffixes are eliminated using these algorithms, which also constantly verify that it is not eliminating any of the roots. The remaining word is then compared against patterns of the same length to extract the root [8]. The following are the three crucial steps that must be taken to establish a classification system [5, 8, 9].
1. Collect and tag the text documents in corpora.
2. Choose a set of features to represent the defined classes.
3. The classification algorithms must be trained and tested using the collected corpora.

Text classification is an important approach in text mining since every unsupervised document on the web seeks to be categorized using a classification method. However, in the Arabic language there is very limited work in automatic classification [10]. Next, some ways for classifications are presented, and some experiments on these ways are also presented.

## 4.1 Classification of Arabic Texts using K-NN and NB Algorithms

The K- Nearest Neighbor (K-NN) and Naïve Bayes (NB) algorithms are the most widely used for categorizing and classifying documents. These algorithms are easily adaptable to work with multiple languages because they do not rely on any particular linguistic rules.

Two pre-processing phases must be completed before the previous classifier can categorize Arabic documents. Stop-word elimination is the first, and the weighting assignment phase is the second.

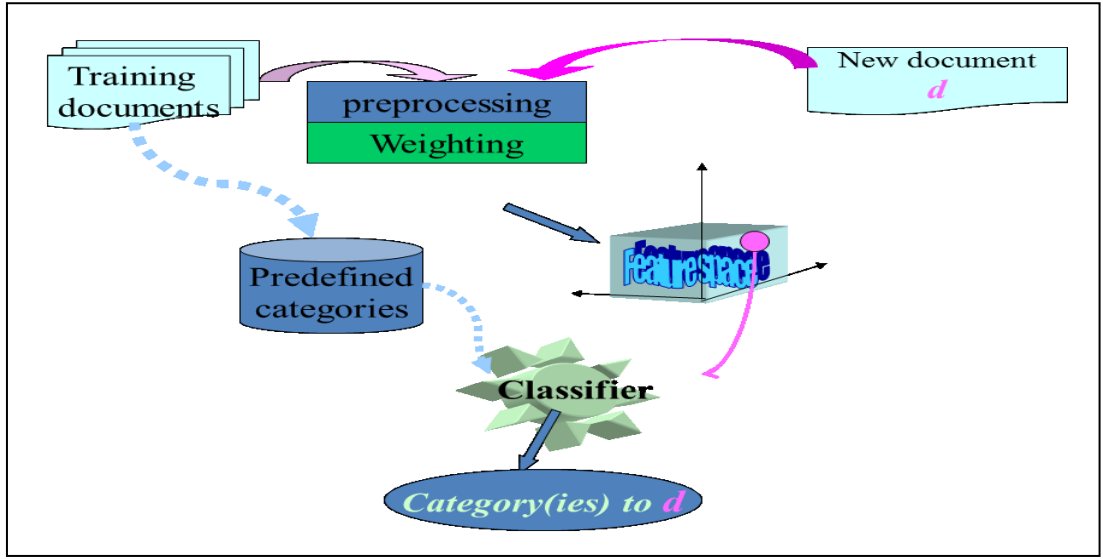

Figure 2: Process of classification algorithm

Stop-word elimination phase is the process of removing the stop words in the document. Stop words are any words, including prepositions, pronouns, and conjunctions like "the," "she," "he," "is," "a," "an," etc., that have no bearing on the body of the sentence [11]. The popular method for getting rid of stop words is based on pre-made lists. While weighting assignment phase is defined as the assignment of a real number between 0 and 1 to each keyword, and this number indicates the imperativeness of the keyword inside the document [12]. To find the weight of each keyword in each document, the document represents as vector space models: $d=(w_1,w_2,\ldots w_t)\in R^t$; where $w_i$ is the weight of ith term in document d. The two most common models for word/term weighting are the Term Frequency weighting scheme (TF) and the Term Frequency-Inverse Document Frequency (TF-IDF) scheme [13, 14]. Below are their explanations, followed by a description of the K-NN and NB algorithms.

- **Term Frequency weighting scheme (TF)**

    $w_{ij} = Freq_{ij}$.

    *Where $Freq_{ij}$ is the number of times jth term occurs in document $D_i$.*

- **Term Frequency - Inverse Document Frequency weighting scheme TF-IDF**

    $w_{ij} = Freq_{ij} * log(N / DocFreq_j)$.

    *Where N =* the number of documents in the training document collection.

*DocFreqj* = the number of documents in which the jth term occurs.

**EXAMPLE: Suppose we have two documents D1, D2 as follow:**

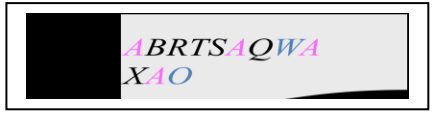 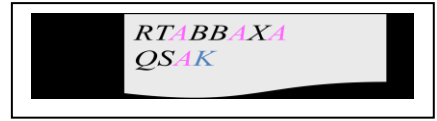

- Using TF, the result will be :

|    | A | B | K | O | Q | R | S | T | W | X |
|----|---|---|---|---|---|---|---|---|---|---|
| D1 | 3 | 1 | 0 | 1 | 1 | 1 | 1 | 1 | 1 | 1 |
| D2 | 3 | 2 | 1 | 0 | 1 | 1 | 1 | 1 | 0 | 1 |

- By TF-IDF , the result will be :

|    | A | B | K   | O   | Q | R | S | T | W   | X |
|----|---|---|-----|-----|---|---|---|---|-----|---|
| D1 | 0 | 0 | 0   | 0.3 | 0 | 0 | 0 | 0 | 0.3 | 0 |
| D2 | 0 | 0 | 0.3 | 0   | 0 | 0 | 0 | 0 | 0   | 0 |

### 4.1.1 Classifier Naive Bayesian (CNB)

Bayesian learning is a probability-driven algorithm based on Bayes probability theorem as follows [15, 16]:

```
1. Input: new document d;
2. Predefined categories: C={c1,c2,….,cl};
3.//Compute the probability that d is in each class c ∈C
   For (ci∈C) {
```
$$\Pr(c_i|d) = \frac{\Pr(d|c_i)\Pr(c_i)}{\Pr(d)} = \prod_i \Pr(word_i \| c_i)\Pr(c_i)$$
```
   }
4.//output: Assigns to d the category c with the highest probability:
```
$$\Pr(d|c) = \max_{\vec{c} \in C} \left(\Pr(d|\vec{c})\right)$$

Where:

**Pr ($c_i$| d)**: It's the probability that a given document d belongs to a given class $C_i$, and that is our target.

**Pr (d):** The probability of a document; we can notice that pr(d) is a constant divider to every calculation, so we can ignore it.

**Pr (ci):** The probability of a class $_i$ (or category); we can compute it from the number of documents in the category divided by documents number in all categories.

**Pr(d | ci )**: It's the probability of a document given class, and documents can be modelled as sets of words thus, the Pr(d |$c_i$ ) can be written like that:

$$Pr(d \mid ci) = \prod_i Pr(word_i \mid c_i)$$

### 4.1.2 Classifier K-Nearest Neighbor (CK-NN)

K-Nearest Neighbor is a widely used text classifier, especially in non-English text mining, because of its simplicity and efficiency [17, 18]. It works as follows:

```
1. Input: new document d;
2. Training collection:D={d1,d2,…dn };
3. Predefined categories:C={c1,c2,….,cl};
4.//Compute similarities
   For (di∈D){   Simil(d,di) =cos(d,di); }
5.//Select k-nearest neighbor
   Construct k-document subset Dk so that
   Simil(d,di) < min(Simil(d,doc) | doc ∈Dk) ∀di ∈D- Dk.
6.//Compute score for each category
   For (ci∈C){
              score (ci) =0;
              For (doc∈Dk) {score (ci) += ((doc∈ci) =true? 1:0)}
              }
7.//Output:Assign to d  the  category c with the highest score:
     score(c) ≥ score (ci), ∀ci ∈C- {c}
```

The work of Zubi [5], who conducted an experiment to classify Arabic texts using the preceding two Algorithms, is presented in this section. He proposed an Arabic text classification system called Arabic Text Classifier (ATC). The primary objective of ATC is to compare the outcomes of the two classifiers that are utilized, CKNN and CNB, and then choose the best average accuracy result to begin the retrieving process. The data used in his work are collected from numerous Arabic websites. The dataset consists of 1562 Arabic documents of different lengths that belongs to 6 categories, the categories are (Economic, Cultural, Political, Social, Sports, and General). This data underwent two pre-processing stages: the first involved deleting stop words and the second included applying TF-IDF to assign weight to each keyword. The experimental findings of this work demonstrated that, with an accuracy of 86.02%, the Classifier K-Nearest Neighbors (CK-NN) outperforms the Classifier Naive Bayesian (CNB). This means that the ATC system will classify and extract Arabic text using the CK-NN rather than CNB.

## 4.2 Classification of Arabic Texts using N-Gram Frequency Statistics

This type uses an N-gram frequency statistics technique that employs a dissimilarity measure, such as the Manhattan distance and Dice's measure of similarity [19, 20]. The descriptions of the N-gram, Manhattan, and Dice measures are provided here.

- **N-gram** is a subsequence of n items from a given sequence. According to the application, this item can be phonemes (smallest segment unit of sound), syllables, letters or words. An N-gram of size one is referred to as a unigram; size two is referred to as a bigram; size three is referred to as a trigram; size four or more is simply called N-gram. It is a type of probabilistic model for predicting the next item in such a sequence. N-grams are used in various statistical natural language processing areas and genetic sequence analysis. For example, the tri-grams for the word المودعين are with N =3:

  الم – لمو – مود – ودع – دعي – عين

- **Manhattan measure**: It calculates a rank-order statistic for two profiles by measuring the difference in the positions of an N-gram in the two different profiles. For each N-gram in the document profile, search for the N-gram in the class profile and calculate the difference in their positions. A maximum value is assigned for N-grams absent from the class profile. After all N-grams in the document profile have been exhausted, the sum of the distance measures is computed. The equation of it as follows:

  $$\text{Manhattan}(P_i, P_j) = {}^k\Sigma_{h=1} |(P_{ih} - P_{jh})|$$

  where $P_i$, $P_j$ represent two N-gram profiles. The class for the document being categorized is the one with the shortest Manhattan distance.

- **Dice's measure:** The Manhattan measure is a dissimilarity metric that compares two profiles. Dice's measure, on the other hand, measures similarity. The class for the text document being categorized is therefore determined by the class having the largest measure.

Khreisat's work in [20] classified Arabic text documents using the N-gram Frequency statistics technique employing a dissimilarity measure called the Manhattan distance and Dice's similarity measure. The classification results using these two measures were compared in terms of recall and precision. This work can be summarised in the following steps:

- A corpus of Arabic text documents was built using Arabic news articles collected from the online websites of several Arabic newspapers. The corpus consisted of text documents covering four categories: sports, economy, technology and weather. The technology and weather documents were very small in size ranging from 1 KB to 4 KB. Sports, and economy documents were much larger ranging from 2 KB to 15 KB for sports documents and 2 KB to 18 KB for economy documents. 40% of the corpus was used for training classes, and the remaining 60% of the corpus was used for classification.
- In text Pre-processing, all documents, whether training documents or documents used for classification, go through a pre-processing phase. It consisted of the following steps :
  1. Convert text files to UTF-8 encoding.

"UTF-8" is a compromise character encoding that can be as compact as Ascii in English or some European languages but also contain any Unicode character with some increase in file size. The eight means it uses 8-bit blocks to represent a character.

2. Remove punctuation marks, diacritics, non-letters, and stop words.
3. Replace initial أ, إ, آ with ا.
4. Replace final ي with ى.

- An N-gram profile was generated for every training document.

Generating the N-gram profile consisted of the following steps:

1. Split the text into tokens composed only of letters. All digits are removed.
2. Compute all possible N-grams for N=3 (Tri-grams)
3. Compute the frequency of occurrence of each N-gram.
4. Sort the N-grams according to their frequencies from most frequent to least frequent.

This gives the N-gram profile for each document. The N-gram profiles were saved in text files.

- N-gram profile was generated as described above for each document to be classified.
- The N-gram profile of each text document (document profile) was compared against the profiles of all documents in the training classes (class profile) using the two measures Manhattan measure and Dice's measure.
- The results obtained using the Manhattan and Dice measures are compared in terms of precision and recall.

Precision = CC / TCF

Recall = CC / TC   Where,

- CC: number of correct categories (classes) found.
- TCF: total number of categories found
- TC: total number of correct categories.

Results of the experiment show that N-gram text classification using the Dice measure outperforms classification using the Manhattan measure.

## 4.3 Classification of Arabic Texts using SVM and C5.0 decision tree algorithm

This type uses a support vector machine or C5.0 decision tree. The Support Vector Machine (SVM) is a pattern recognition and data analysis method. It serves as a classification tool [16, 21]. While a C5.0 decision tree is a tree in which each leaf node represents a classification or decision and each branch node offers a choice between several alternatives [22, 23]. Al-Harbi et al. [8] implemented a tool for Arabic Text Classification (ATC) to accomplish feature extraction and selection. This tool can perform the following main functions:

1. Automatically divide the dataset into two partitions - training and testing – according to the user input of training and testing size.

2. Extract the lexical features (single word) and generate the feature frequency profile for both the training set and testing set with options to explore each class and file profile.
3. Calculate the importance of each feature locally (for each class) based on Chi-Square statistics.
4. Generate training and testing matrices.

They used a corpus comprised of 17,658 texts with more than 11,500,000 words. Such diversity in genre, class and text length will provide a clear insight into the classification algorithms' performance and their ability to classify Arabic texts.

Table 1: The number of text in each used dataset.

| Genre | No. of Text | Classes |
| --- | --- | --- |
| **Saudi Press Agency (SPA)** | 1,526 | Cultural News, Sports News, Social News, Economic News, Political News, General News |
| **Saudi News Papers (SNP)** | 4,842 | Cultural News, Sports News, Social News, Economic News, Political News, General News, IT News |
| **WEB Sites** | 2,170 | IT, Economics, Religion, NEWS Medical, Cultural, Scientific |
| **Writers** | 821 | Ten writers |
| **Discussion Forums** | 4,107 | IT, Economics, Religion, NEWS Medical, Cultural,Scientific |
| **Islamic Topics** | 2,243 | Hadeeth, Aqeedah, Lughah, Tafseer, Feqah |
| **Arabic Poems** | 1,949 | Hekmah, Retha, Ghazal, Madeh, Heja, Wasf |
| **Total** | 17,658 | |

Their method involves applying the Chi-Squared statistic to select the training dataset's top 30 terms from each class. Where Chi-Squared statistic ($X^2$) computed the dependency between two factors: the term t and the class c, and considered as a test with one degree of freedom. The higher the value of Chi-Squared, the higher the dependency or association between the term and the class. The following table of a term t and the class c can be used to illustrate the idea.

Table 2: The contingency table of t and c

|  | c | Not c | Total |
| --- | --- | --- | --- |
| **t** | A | B | A+B |
| **Not t** | C | D | C+D |
| **Total** | A+C | B+D | N |

The following equation illustrates Chi-Squared statistics using Table 2.

$$X^2(t, c) = \frac{N(AD-CB)^2}{(A+C)(B+D)(A+B)(C+D)}$$

Where N=Total number of documents in the corpus.
A= Number of documents in class c containing the term t.

B= Number of documents that contain the term t in other classes.

C= Number of documents in class c that do not contain the term t.

D= Number of documents that do not contain the term t in other classes.

Their experiment aims to evaluate the performance of two popular classification algorithms (SVM and C5.0 decision tree algorithm) on classifying Arabic text using the seven Arabic corpora described above. The result shows that the C5.0 algorithm outperformed the SVM algorithm by about 10%.

## 4.4 Classification of Arabic Texts using Deep learning algorithm

Deep learning (DL) approaches have recently received particular attention in natural language processes such as text mining and opinion mining [24, 25]. There have been numerous efforts to classify Arabic texts using deep learning, particularly for the Arabic language, such as [26-28]. Elnagar et al. [27] compared multiple DL models for categorizing Arabic text to evaluate how well different DL models for classifying Arabic text perform on their collected corpus. They use a different strategy from others in that they skip the pre-processing stage. Findings showed that the attention-gated recurrent unit (GRU) model achieves the highest accuracy compared to the other used models. Another work proposed by Boukil et al. [28] utilized an Arabic stemming algorithm to extract and reduce the required features. Then, as a feature weighting technique, they employed the TF-IDF method. Finally, they used the convolutional neural networks (CNN) algorithm for the classification stage. Results demonstrated that their method could produce outstanding results on various benchmarks.

Moreover, Alsaleh et al. [29] proposed a hybrid classification model utilizing CNN and genetic algorithms. Their model aims to overcome the drawback of current studies that have failed to achieve high classification accuracy due to parameter setting issues by using genetic algorithm that is capable to optimize the CNN parameters. Using two large datasets, the proposed model is evaluated and contrasted with state-of-the art studies. The outcomes demonstrated a 4–5% improvement in the categorization accuracy. Finally, there are some other studies in this field and the readers who are interested can have a look to the following studies:

- Sundus et al. [30] introduced a supervised feed forward Deep learning, and their findings showed that the Arabic text classification issue is up-and-coming with deep learning classification models.
- Alhawarat and Aseeri [31] introduced a Superior Arabic Text Categorization Deep Model, SATCDM, using a CNN multi-kernel architecture with word embedding and specifically n-gram to categorize Arabic news documents. Their approach achieves very high precision using 15 of the publicly available datasets. Their model used 15 publicly accessible datasets and achieved high accuracy.

## 5. New Arabic Classification System

The proposed classification system collects Arabic documents from two newspapers, Al-Watan Newspaper and Al-Namr Sports Newspaper. Each Arabic document is copied individually from the Html page to the text file before being encoded with UTF-8. All the documents which are collected are put in a corpus. Some of these documents are used as training documents to identify our corpus's available classes, and others are used as testing documents. Our system has two categories (classes): sports and politics. Each training document goes through these steps:

**Step 1.** Divide the text into Tokens.
**Step 2.** Delete repeated words.
**Step 3.** Delete Empty tokens, numbers and critics marks.
Remove Critics means change the form of some letters as follow:

آ – إ – أ → ا   &&   لا - لأ – لإ → لا   &&   ؤ → و

ى → ي - ئ   &&   ة → ه

**Step 4.** Every token goes through the Light stemmer function, which is responsible for deleting prefixes and suffixes from the word. As mentioned in phase 1, this function does well in some cases and fails in some words.
**Step 5.** Insert the tokens of the first document of the first class into a file which will be considered as the class "Sport Class or Politics Class".
**Step 6.** After implementing Step5 for the first documents in sports documents and for the first document in politics documents, we will have both sports file (will be the sport class) containing all the tokens in the first sport document and the politics file (will be the politics class) contains all the tokens in the first politics document.
  The next document of class n "either sports or politics" goes through the first three steps, and then all the tokens of the new document are compared with all the words in the class file "if its sports doc is compared with the sport class otherwise it will be compared with the politics class". The new tokens which are founded in the new documents and not found in the class file are only added to the class file; the others will be discarded.
**Step 7.** Step6 is repeated for every training document until they all have been processed.
**Step 8.** The Sport Class file will be generated, which contains most of the words that may appear in any sport document. The Politics Class file will also be generated, which contains most of the words that may appear in any politics document.

Any document 'd' from testing documents goes through the first three steps, then the Classifier Naive Bayesian (CNB), which describe in section 4.1.2, is applied to determine the probability of belonging document d with all the available classes and choose a class for the document which has the highest probability.

All the details of the system regarding experimental results and evaluation are discussed in [32].

## 6. Conclusion and Future works

Text mining has become a key area of research due to the abundance of Arabic texts currently available on the internet and the desire for a better method of accessing and retrieving this data. Various studies are exploring different ways to classify documents written in various languages. Due to the intricacy of the Arabic language and its nature, there is not much ongoing study on Arabic document classification. This work presents some algorithms used for Arabic document classification, starting with machine learning to deep learning algorithms. A new Arabic categorization system is then proposed to categorize Arabic text documents in order to enhance analysis and facilitate text retrieval for Arabic texts. The new Arabic classifier is tested on many Arabic texts, effectively identifying the correct class for every test text. In our future work, we intend to broaden the system's capabilities and enable it to identify any class (such as Politics, Islam, Economics, News, etc.) while utilizing deep learning techniques for classification.